\documentstyle[draft,12pt]{article}
\newcommand{\be}{\begin{equation}}
\newcommand{\ee}{\end{equation}}

\begin{document}

\title{The Phase Structure of Strings with Extrinsic Curvature
}
\author{
Mark Bowick\thanks{Invited talk delivered by Mark Bowick at the
Workshop ``String Quantum Gravity and Physics at the Planck Energy
Scale," Erice, June 21-28, 1992 and at the 1st Iberian Meeting on Gravity,
\'Evora, Portugal, September 21-26, 1992.},
Paul Coddington, Leping Han,\\
Geoffrey Harris and Enzo Marinari$^{(\dag)}$\\[1.5em]
Dept. of Physics and NPAC,\\
Syracuse University,\\
Syracuse, NY 13244, USA\\
{\footnotesize
  bowick@suhep.phy.syr.edu
  paulc@sccs.syr.edu
  han@sccs.syr.edu}\\
{\footnotesize
  gharris@sccs.syr.edu
  marinari@roma2.infn.it}\\[1.0em]
{\small $(\dag)$:  and Dipartimento di Fisica and INFN,} \\
{\small Universit\`a di Roma {\it Tor Vergata}}\\
{\small Viale della Ricerca Scientifica, 00173 Roma, Italy}}
\maketitle
\begin{flushright}
  {\bf SU-HEP-4241-526}\\
  {\bf hep-th/9211058}\\
\end{flushright}
\pagebreak
\begin{abstract}
We examine a model of non-self-avoiding, fluctuating surfaces as a candidate
continuum string theory of surfaces in three dimensions.  This model describes
Dynamically Triangulated Random Surfaces embedded in three dimensions with an
extrinsic curvature dependent action.  We analyze, using Monte Carlo
simulations,
the dramatic crossover behaviour of several observables which characterize the
geometry
of these surfaces.  We then critically discuss whether our observations
are indicative
of a phase transition.
\end{abstract}
\pagebreak

\section{Introduction}

In this work, we investigate a theory of fluid,
fluctuating random surfaces embedded
in three dimensions.  Various theories of fluctuating surfaces (string
theories)
arise in the description of many physical systems.  Among these are
the 3-dimensional Ising model, the strong interaction (QCD) in the infrared
limit and fundamental unified theories of all interactions including
gravity.
Natural biological membranes, such as lipid bilayers, and artificial membranes,
such as micelles and vesicles, also form a rich class of fluctuating
surfaces together with interfaces, such as those in microemulsions,
between two distinct three-dimensional bulk phases \cite{DAVID}.  These
latter systems are fluid because their
component `molecules' are loosely bound.  Their constituents are
arranged so that their net surface tension nearly vanishes; thus these
membranes undergo large thermal fluctuations.  These
biological and chemical membranes exhibit self-avoidance,
which we do not take into account in our simulations.

Just as field theories are described by sums over paths, string theories
are formally characterized by a functional integral $Z$
which can be written as a sum over surfaces weighted by $\exp (-S)$.
Here $S$ denotes the action associated with a particular surface. Note that
we work in Euclidean space.
To write down our action we introduce an explicit parametrization of a
generic surface $\cal M$ in $R^{3}$ with coordinates $(\sigma _1, \sigma _2)$
and the embedding $X^{\mu}(\sigma _i)$. $\mu$ runs from $1$ to $3$ (since
we only study the case of a $3d$ embedding space).  The induced metric
(the pullback of the  Euclidean $R^{3}$ metric via the embedding) is given by

\be
  h_{ij} = \partial _{\sigma_i} X^{\mu}\partial _{\sigma_j} X_{\mu}\ .
\ee

We will use Greek letters for the embedding space indices; Roman
letters label the coordinate basis on each surface.  A particularly
natural choice for the action is the Nambu-Goto action, which is proportional
to the area of the surface in the embedding space, and is given by
\be
S = \mu\int{\sqrt{{\rm{det}}(h)}}.
\ee
This action is difficult to quantize, though, since it is non-polynomial.
A further complication is that the measure in the path-integral
constructed with this action is quite
subtle, especially  when the theory is discretized.  By
introducing additional degrees of freedom, in the form of an
intrinsic metric $g_{ij}$ on each surface, one can write down the
Polyakov action \footnote{An additional term, which is a topological
invariant proportional to the Euler
character of the surface is added to both the Polyakov and Nambu-Goto
actions.  The coefficient of this term is referred to as the `string
coupling'. It is the only dimensionless parameter in the theory; therefore
it serves as the perturbative expansion parameter for string theory.}.
\be
  S = \int\sqrt{|{\det {g}}|}(g^{ij}\partial_i X^{\mu}
  \partial_j X^{\mu}).
\ee
This action has proven to be much more tractable analytically.  Note
that it possesses a large degree of gauge freedom, associated with
its invariance under reparametrizations of the metric and the intrinsic
surface coordinates.  Often, one gauge-fixes the metric to the form
$g_{ij} = \exp(\rho)\delta_{ij}$; $\rho$ is referred to as the Liouville
mode.  The Polyakov action is in fact independent of $\rho$
\footnote
{It follows that substituting the solution to the equations of
motion back into the Polyakov action yields the
Nambu-Goto action.  At the quantum level, though, it has only been
shown that these actions are equivalent for $D=26$\cite{MORRIS}.  In lower
dimensions this equivalence has been questioned, for example,
by Distler \cite{DIST3D}.  The work of Polchinski and Strominger
also suggests that there are alternate quantizations \cite{POLSTRO}.}
, but this
`Weyl' symmetry is anomalous in general.  The anomaly is expressed in
terms of the central charge $c$.  By carefully gauge-fixing the entire
path-integral one can derive the anomaly condition
\be
      c_{\rm{matter}} + c_{\rho} - 26 = 0.
\ee
The first term depends on the space in which the surfaces are embedded,
or, in other words, on the particular field theory that lives on
the two-dimensional surface.  For instance, the central charge associated
with $D$ bosons is $D$ ($1$ per boson); this characterizes a string
embedded in $R^{D}$ \footnote{ A fermion has a central charge of $1/2$,
so that by studying an Ising model on random fluctuating surfaces, we
are effectively studying strings imbedded in `fractional' dimensions.}
The $-26$ arises from the anomaly contribution from ghosts needed
for gauge-fixing reparametrization invariance.  Note that when $D=26$,
$c_{\rho}$, the Liouville central charge, vanishes and the Liouville
mode effectively decouples from the theory.  For $D<26$ one can integrate
the matter ($X^{\mu}$) fields out and obtain an effective Liouville theory.
The Liouville theory is highly non-linear; furthermore, its analysis is
complicated by the fact that
the two-dimensional metric on the surface in which the Liouville
field lives depends on the field itself!  Nevertheless, with a few general
assumptions about the path integral measure,
the spectrum of the theory at each order in perturbation
theory can be computed when $c_{\rm{matter}} \leq 1$.  Additional exponents,
such as the Hausdorff dimension, which characterize both the intrinsic
and extrinsic geometry of these surfaces, have also been computed
from Liouville theory.  These theories are often analyzed on surfaces
of fixed area; in this case, the string susceptibility $\gamma_h$
is determined by the dependence of the generating functional for surfaces
of genus $h$ on the
area $A$
\be
\ln(Z_h) = \exp(\mu_cA)A^{\gamma_h -3},
\ee
where $\mu_c$ is a non-universal constant.  An analysis of the Liouville model
yields
\be
\label{gammas}
\gamma_h = 2 - \frac{(1-h)}{12}(25 - c_{\rm{matter}} +
\sqrt{(1-c_{\rm{matter}})(25 - c_{\rm{matter}})}).
\ee
These computations
break down in the regime $1 < c_{\rm{matter}} < 25$, where exponents
such as the string susceptibility given in (\ref{gammas}) become imaginary.

  Further progress in understanding string theory in a low number of dimensions
has been made by mapping the sum over surfaces onto an integral over
matrices.  This mapping is realized by  replacing continuum surfaces
by their discrete cellular decompositions, such as their triangulations.  In
this construction,
each triangle is equilateral with area $1$ in the intrinsic  metric; the
coordination number at each vertex determines the intrinsic curvature of the
surface.  The coordinates $i$ label the vertices of the triangulation. Then
the discrete analogue of the intrinsic metric is the adjacency matrix
$C_{ij}$ whose elements equal $1$ if $i$ and $j$ label neighbouring nodes of
the triangulation, and vanish otherwise.  Two-dimensional diffeomorphism
invariance reduces, at this discrete level, to the permutation symmetry
of the adjacency matrix.  One
of the keys, in fact, to the power of this
construction is the preservation of this symmetry.
The triangulation of a surface of genus $h$ is then dual to a phi-cubed
diagram of genus $h$.  The large
$N$ ('t Hooft) expansion of an integral over $N \times N$ Hermitian matrices
generates these Feynman graphs.  In this case the area of the surfaces
is not fixed, since all graphs of fixed genus are summed over.  The action
then acquires a contribution (from Legendre transformation) proportional
to the product of the cosmological constant with surface area $A$.
Orthogonal polynomial techniques can
then be applied to analyze the matrix integral. Indeed the theory
can be exactly solved in the `double scaling limit' \cite{EXACT} in which the
string
coupling, cosmological
constant and matter couplings (which are determined by the matrix
`potential') are tuned together.  These solutions are exact,
and thus include non-perturbative information about these string theories,
albeit with ambiguities in certain cases.  Unfortunately, the matrix models
that represent theories with $c_{\rm{matter}} > 1$ are too difficult
to solve exactly.

    It has generally been suspected that these analytic techniques fail
for $c_{\rm{matter}} > 1$
because the string theory becomes pathological.  In this
regime, a negative mass-squared
particle, which is sometimes referred to as the
`tachyon', comes on shell \footnote{More
precisely, one encounters these instabilities in Liouville theory when the
quantity $c - 24\Delta  >1$, where $c$ denotes the central charge of the matter
theory which describes the embedding of the surfaces and $\Delta$ is the
conformal weight of the lowest weight state in this theory \cite{KUTSEI}.
Since here we are considering flat space, $c=D$ and $\Delta=0$.}. It
is thought that its presence might
engender the proliferation  of long tubes with thickness of the scale
of the ultraviolet cutoff on surfaces that dominate the string
functional integral.  Cates, by analyzing the Liouville action,
has put forth another perhaps related explanation of this pathology
\cite{CATES}.
He points out that in the presence of vortex configurations of the form
$\rho \sim -\ln(r)$, reducing the world-sheet cutoff to zero no longer
causes the spacetime cutoff to vanish.  He then shows that
these vortices have negative free-energy when $c_{\rm{matter}}  > 1$
and proliferate to the extent that the Liouville partition function
becomes ill-defined with a finite spacetime cutoff and zero
cosmological constant.  For positive cosmological constant, it then
appears plausible that these vortex configurations would still appear,
though perhaps the partition function will be finite.  The
predominant surfaces in the functional integral would thus be subject to
large fluctuations of their internal geometry.
Recent simulations of
multiple Potts models with $c_{\rm{matter}}$ less than, equal to,
and greater than $1$ coupled to gravity have been performed \cite{MPOTTS}.
These
simulations, however, do not show any dramatic changes in
the behaviour of the internal geometry of the dominant surfaces
as $c_{\rm{matter}}$ is increased beyond $1$.  More analytic and numerical
work clearly needs to be done to determine what exactly happens
as $c_{\rm{matter}}$ becomes greater than $1$.

   Monte Carlo simulations of strings embedded in flat space for
$D > 1$
do indicate that these theories fail to describe the fluctuations of
two dimensional {\em smoothly embedded} surfaces in the continuum limit.
The normals to the surfaces dominating the simulations are
uncorrelated over the distance of a few lattice spacings.  These
surfaces also appear to have a large (greater than $8$), or
perhaps infinite, extrinsic Hausdorff dimension; they resemble
branched polymers \footnote{As above, singular configurations
also dominate the Gaussian theory, which is essentially a theory
of free random walks, rather than surfaces.}.  Such configurations should
not describe, for instance, the domains of the 3d Ising model or QCD strings.
We would thus like to find a modified string theory that is dominated by
smoother surfaces.

The tachyon, and apparently
these related instabilities, can be eliminated in
particular cases by introducing fermionic coordinates and supersymmetry on
the worldsheet, and implementing an appropriate projection of states.
The fermions, presumably, effectively smooth out the
surfaces.  This would be consistent with what has been observed for
one-dimensional geometries; the random walk of a spin one-half particle has
Hausdorff dimension one and thus appears to be smooth \cite{POLBOO}. Many
authors have proposed an alternative modification of the string action
\cite{CANHAM,HELFRI,POLEXT,KLEINA} via the addition of a term that directly
suppresses extrinsic curvature\footnote{In fact, integrating the fermions out
of the Green-Schwarz superstring yields an action similar to the one we
consider, but with the addition of a complex Wess-Zumino type term
\cite{WIEGMA}.}. We shall examine this class of theories in this talk.

   To characterize the geometry of our surfaces further, we associate
with each
point in our generic surface $\cal{M}$  tangent vectors  ($t^{\mu}_i {\in}
 T\cal{M}$) and a normal
vector  $n^{\mu} {\in} T\cal{M}^{\perp}$.
The extrinsic curvature matrix $K_{ij}$
(the second fundamental form) can be defined by

\begin{equation}
\protect\label{kijdef}
         \partial_{i} n^{\mu} = - K_{ij}{t^{\mu}}^j \ .
\end{equation}
The eigenvalues of this matrix are the inverses of the radii of curvature
of $\cal{M}$. One usually describes the geometry of
these surfaces in terms of the mean curvature \cite{CHOBRU,DOCARM}

\begin{equation}
            H = \frac{1}{2}h^{ij}K_{ij}\ ,
\end{equation}
and the Gaussian curvature

\begin{equation}
           K = \epsilon^{ik}\epsilon^{jl}K_{ij}K_{kl}\ .
\end{equation}

One can show that the Gaussian curvature can be computed solely from
the metric $h_{ij}$, while the mean curvature depends explicitly on
the embedding $X^{\mu}$.

    Our lattice model is constructed by triangulating each surface, as
we discussed above in the context of matrix models.
Each node of the triangulation is embedded in $R^{3}$ by the functions
$X_i^{\mu}$; $i$ labels the $i$th node and $\mu$ runs from $1$ to $3$.
We also associate a normal
vector $(n^{\mu})_{\hat{k}}$ with each triangle (indices with hats
label the triangles).  We shall study the theory defined by the action
\begin{equation}
\protect\label{ourdisact}
          S =  S_{G} + \lambda S_E =
    \sum_{i,j,\mu}C_{ij}(X^{\mu}_i - X^{\mu}_j)^2 +
 \lambda\sum_{\hat{k},\hat
{l},\mu}C^{\hat{k}\hat{l}}(1 -
  n^{\mu}_{\hat{k}}\cdot n^{\mu}_{\hat{l}}) .
\end{equation}
This model has been examined in references \cite{CAT,BJW,BCJW,CKR,A1,A2,US}.
For $\lambda = 0$ this is simply a discretization of the Polyakov string
action.  The final term, which depends on the discretized extrinsic
curvature, introduces a ferromagnetic interaction
between surface normals, which one might hope would cause smoother
surfaces to dominate the partition function.
We would like
to know if there is a smooth phase and a phase transition (at
some finite $\lambda_c$) between this phase and the
crumpled phase observed
at $\lambda = 0$.  If this were so, an interesting
continuum limit of this lattice model could perhaps be constructed
at this phase transition point, yielding a new continuum string theory.

   The action we simulate is in fact a particularly natural discretization
of
\begin{equation}
  \protect\label{ourcontact}
  S = \int\sqrt{|{\det {g}}|}(g^{ij}\partial_i X^{\mu}
  \partial_j X^{\mu}  + \frac{\lambda}{2}g^{ij}h^{kl}K_{ik}K_{jl})\ .
\end{equation}
Note that the
second term in the action is manifestly positive, Weyl and reparametrization
invariant, and that $\lambda$ is a dimensionless coupling.  So,
it is not clear whether or not this term is relevant.  Mean field (large $D$)
and perturbative RG calculations have been performed using
similar actions, such as
\begin{equation}
  \protect\label{acthkf}
     S = \int d^2\sigma\ (\mu_0\sqrt{\det{h}} + \frac{1}{\alpha}
    \sqrt{\det{h}}(h^{ij}K_{ij})^2)\ ,
\end{equation}
in the regime in which the string tension $\mu_0$ is small (unlike the usual
particle physics limit of string theory, which is characterized by large
$\mu_0$). After integrating out fluctuations of the embedding $X^{\mu}$
between momentum scales $\Lambda$ and $\tilde{\Lambda}$, it is found that the
renormalization of the extrinsic curvature coupling is given to one-loop
order by

\begin{equation}
  \beta ({\alpha}) \equiv \Lambda \frac{d\alpha}{d\Lambda}
  = - \frac{3}{4\pi}\alpha ^2 \ ,
\end{equation}
so that $\alpha$ is driven to infinity in the infra-red.
This theory thus exhibits asymptotic freedom. Surfaces are smooth
(the normals are correlated) below a persistence length \cite{PELLEI}

\begin{equation}
  \protect\label{persist}
  \xi_p \sim \exp(\frac{4\pi}{3\alpha_{bare}})\
\end{equation}
and are disordered above this scale.  Some
intuition into this result can be gained by observing that this theory is
similar to the $O(3)$ sigma model, which is  asymptotically free
\cite{POLBOO}. The normals to $M$ are the analogues of $O(3)$ vectors, though
in this case they are constrained to be normal to a surface governed by the
action (\ref{ourcontact}).

    The analytic results, therefore, do not indicate that we should anticipate
a finite $\lambda$ phase transition.  Note, though, that the RG calculations
are based on a Nambu-Goto type action although we simulate an extension of
the Polyakov action.  Since the two actions are not clearly equivalent,
{\em particularly} when extrinsic curvature dependent terms are added, we
cannot simply assume that these analytic and our numerical results should
agree.

\section{The Simulation}
     We have considered triangulations with the topology of the torus,
to minimize finite size effects.
The above action  was used, with
the BRST invariant measure utilized also by Baillie, Johnston and Williams
\cite{BJW}, so that

\begin{equation}
  Z = \sum_{G \in T(1)}
  \int\prod_{\mu,i}dX^{\mu}_{i}\prod_{i}q_i^{\frac{D}{2}}
  \exp( -S_{G} - \lambda S_E)\ ,
\end{equation}
where $D=3$, $q_i$ is the connectivity of the $i$th vertex and $T(1)$ refers
to the set of triangulations of genus $1$.  We used the standard Metropolis
algorithm to update our configurations.
To sweep through the space of
triangulations we performed flips on randomly
chosen links.  Flips were automatically rejected if they yielded a degenerate
triangulation.
After a set of  $3M$ flips was
performed, $3M$ randomly selected embedding coordinates were updated via
random shifts from a flat distribution.
Most of the  Monte Carlo simulations were performed on
HP-9000 (720 and 750 series) workstations; we also collected some data by
simulating lattices on each of the 32 nodes of a CM-5.


%

We ran on lattices ranging in size from $N = 36$ to $576$ ($N$ signifies
the number of vertices) with $4$ to $7$ different values
of $\lambda$ for each $N$.  Most of the data was this data was taken in the
region $\lambda \in (1.325,1.475)$.  For small $N$, the runs consisted
of $3 \times 10^6$ sweeps, while we performed longer runs (of up to
$27 \times 10^6$ sweeps for $N=576$) for larger lattices,
because the auto-correlation
times for our simulations were very large.  (The correlation time
for the radius of gyration was greater than $10^6$ sweeps
for $N=576$!)  To determine our
observables as a function of $\lambda$ we used a histogram
reconstruction procedure.  We patched different histograms by
weighting them with the associated statistical indetermination
(which was estimated by a jack-knife binned procedure).  Various consistency
checks indicate that this procedure is very reliable.
\section{Observables}
     We measured the edge action $S_E$ and the associated specific heat
$C(\lambda) \equiv \frac{\lambda^2}{N}(<S_E^2> - <S_E>^2)$.   In Fig. 1
we plot the specific heat curve (constructed via the histogram procedure)
and we tabulate its maximum and peak position for various lattice sizes
in Table $1$.
\begin{table}
\begin{tabular}{|l|l|l|} \hline
N & $C(\lambda)^{\rm{max}}$ & $\lambda_c$ \\ \hline
36 & 3.484(8) & 1.425(35) \\ \hline
72 & 4.571(15) & 1.410(15) \\ \hline
144 & 5.37(14) & 1.395(30) \\ \hline
288 & 5.55(7) & 1.410(25) \\ \hline
576 & 5.81(17) & 1.425(30) \\ \hline
\end{tabular}
\protect\caption[CT_TWO]{The maximum of the specific heat and its position,
with errors, for different lattice sizes. \protect\label{T_TWO}}
\end{table}

     We see that the specific heat peak grows vigorously with $N$ for small
lattices, but that this growth quickly levels off for larger $N$. These
observations agree fairly well with previous work \cite{CKR,A1,A2}. For the
larger lattices it appears that the peak position shifts very slowly
towards higher values of $\lambda$, though this increase is not
statistically significant.  The shape of the peak does not change dramatically
with $N$; it narrows perhaps a bit between $N=144$ and $N=576$.

     To determine how the mean size of the dominant surfaces depend on
$\lambda$,
we measured
the squared radius of gyration $R_G$;
  \begin{equation}
    R_G \equiv \frac{1}{N}\sum_{i,\mu}(X_i^{\mu} - X_{{\rm{com}}}^{\mu})^2\ ,
    \protect\label{E_RG}
  \end{equation}
where the com subscript refers to the center of mass of the surface.
By measuring the $N$ dependence of the gyration radius, one can extract
a value for the extrinsic Hausdorff dimension, which is given by
  \begin{equation}
    R_G \sim N^{\nu}\sim N^{\frac{2}{d_{\rm{extr}}}}\ \ .
    \protect\label{E_RGD}
  \end{equation}
We plot $R_G$ in Fig. 2(a), clearly the size of the mean surface size increases
dramatically for $\lambda$ near $1.4$.  In Fig. 2(b), we plot
the effective Hausdorff dimension, given by

\begin{equation}
  \nu(N) \equiv \frac{\log \frac{R_G(N)}{R_G(\frac{N}{2})}} {\log(2)}\ .
\end{equation}

In the large $\lambda$ limit $\nu \to 1$ and $d_{extr}\to 2$, as
expected for flat surfaces. In the low $\lambda$ limit $d_{extr}$ becomes
very large. In the pseudo-critical region $\nu$ is a linear function of
$\lambda$. Curiously enough, the latter curve yields a Hausdorff
dimension of $4$, a value characteristic of branched polymers, near the
location of the specific heat peak.
This value is not particularly reliable though because of finite-size effects
and also because it changes rapidly in this region.  In
ref. \cite{A2} a value compatible with ours ($D_H(\lambda_c)>3.4$)
is quoted for the critical theory.
We stress however (and also here we are in complete agreement with
\cite{A2}) that the Hausdorff dimension in the pseudo-critical
region depends heavily and quite unusually on $N$.

In both the high and low $\lambda$ regions finite size effects are
quite small (compatible with zero to one standard deviation).
In the pseudo-critical region, on the contrary,
finite size effects are large. This effect
cannot be explained by the shift in $\lambda$ which one gets from the shift of
the peak of the specific heat, which is far too small.

     We also measured the magnitude of the extrinsic Gaussian curvature,
$\int\mid K\mid\sqrt{\mid h \mid}$, given by
  \begin{equation}
    \mid {\cal{K}} \mid = \frac{1}{N}\sum_i\mid 2\pi -
    \sum_{\hat{j}} \phi_i^{\hat{j}}\mid\ .
    \protect\label{E_CK}
  \end{equation}
Here $\phi_i^{\hat{j}}$ denotes the angle subtended by the $\hat{j}$th
triangle at the $i$th vertex.  This quantity, plotted in Fig. 3,
measures the
magnitude of the deficit angle in the embedding space averaged over
all vertices.
Note that the mean Gaussian curvature decreases rapidly
in the neighborhood of $\lambda = 1.4$, indicating that a sharp
crossover is occuring in this system.  From this plot we can see that
finite size effects increase with $\lambda$.  They do not appear
to peak in the region about $\lambda = 1.4$ as one might expect
for a typical phase transition.

     The magnitude of the intrinsic Gaussian curvature,
$\mid {\cal{R}} \mid$, given by
  \begin{equation}
    \mid {\cal{R}} \mid = \frac{\pi}{3N}\sum_i\mid 6 - q_i \mid\ ,
    \protect\label{E_CR}
  \end{equation}
is shown in Fig. 4.  When the intrinsic and extrinsic metrics are equal,
the intrinsic and extrinsic deficit angles are identical and
$K = R/2$.  Both extrinsic and extrinsic curvatures behave in a qualitatively
similar manner; $\mid {\cal{R}} \mid$ drops off rapidly, just as $\mid
{\cal{K}} \mid$
does.
Through the peak region, though, $\mid {\cal{K}} \mid$ decreases
by about a factor
of $5$ while $\mid {\cal{R}} \mid$ diminishes to only about $.6$ of its value
on the left-hand side of the peak.  Since the action explicitly suppresses
mean curvature, and the mean and extrinsic
Gaussian curvature are closely related
(for instance, $H^2 > \frac{K}{2}$), we would expect that for large
$\lambda$  extrinsic fluctuations would be suppressed much more than
fluctuations of intrinsic geometry.

       The question of the equivalence of the Nambu-Goto and Polyakov
actions motivated us to study the correlations between intrinsic geometry
(which is not introduced independently in the Nambu-Goto formulation) and
extrinsic geometry.  We measured the quantity which we refer to as
${\cal{K}}*{\cal{R}}$

  \begin{equation}
    {\cal{K}}*{\cal{R}} \equiv \frac{\int KR}{\sqrt{\int K^2 \int R^2}} =
    \frac{\sum_i(2\pi - \sum_{\hat{j}}\phi_i^{\hat{j}})(6 - q_i)}{\sqrt{
    \sum_i(2\pi - \sum_{\hat{j}}\phi_i^{\hat{j}})^2\sum_i(6 - q_i)^2}}\ .
    \protect\label{E_KR}
  \end{equation}

  This quantity is $1$ when the metrics are equal, $0$ if they are un-
  correlated, and negative when these curvatures are anti-correlated.
The plot of
${\cal K}*{\cal R}$ in Fig. 5 indicates that intrinsic and extrinsic geometry
are
strongly correlated for small $\lambda$, but as one passes through the peak
region they become decorrelated.  This
is not particularly surprising, given
that the action directly suppresses only extrinsic fluctuations.
Note that RG calculations based on the Nambu-Goto action plus an
extrinsic curvature term
perturb about a background that is both intrinsically and extrinsically
flat.  Given the observed decorrelation
between intrinsic and extrinsic geometry,
we would not anticipate that this background appears in the low-
temperature limit of the model which we simulate.

We also measured various other observables which
characterize both the intrinsic and extrinsic geometry of these surfaces. These
measurements are discussed in another write-up of this work
\cite {US}.  They all exhibit sharp crossover behaviour in
the region near $\lambda= 1.4$.
We found that the auto-correlation times
of these observables grew rapidly as $\lambda$ increased, but we did not
note any maximum in these times in the region about $\lambda = 1.4$.

  The crossover behaviour became also quite apparent
when we examined typical snapshots of our simulated surfaces for various values
of $\lambda$.  In Figs. 6 (a)-(d) we present pictures of typical surfaces for
$\lambda =
.8,1.3,1.5,$ and $2.0$.
Note that the surfaces rapidly become smoother and the normal-normal
correlation length increases significantly as one passes from the second to
the third of these pictures, which correspond to only slightly different values
of
$\lambda$.

\section{Interpretation}
This model of crumpled surfaces appears to
exhibit sharp crossover behaviour in the region around $\lambda = 1.4$.  The
sharp change in the magnitude of the Gaussian curvature,
the radius of gyration and other observables
indicates that the normals acquire long-range correlations, up to
the size of the systems we examine.  The zero string tension
measurement of
\cite{A2} also shows that the disordered regime differs from the regime in
which the surfaces are ordered (up to scale of the lattices that are
simulated) by only  a small shift in $\lambda$. This evidence  might indicate
the presence of a phase transition at this point.  Since the peak
growth rapidly diminishes for large $N$, such a phase transition would
likely be higher than second order.
Still, the apparent absence of
diverging correlation times and, in some cases, increasing finite size effects
in the
peak region leads us to question whether we are actually observing
a typical phase transition.

There are indeed other possible interpretations of our data.
Note that the surfaces which we
simulate are quite small.
For instance, if the surfaces in our simulations had an
intrinsic dimension of $2.87$ (characteristic of $D=0$ gravity),
they would have roughly a linear size of fewer than $10$
lattice spacings
\footnote{Of course, our lattices are too small,  by one or
two orders of magnitude, to really exhibit a convincing fractal structure.}.

Perhaps the simplest alternative explanation  for the presence of this peak is
suggested by the arguments of Kroll and Gompper \cite{KROGOM}.  They argue
that the peak occurs when the persistence length of the system approaches the
size of the lattice ($\xi_p \sim N^{\frac{1}{d}}$); $d$ denotes the intrinsic
Hausdorff dimension.
Fluctuations on a larger scale become more
important.  When this scale is greater than the lattice size these
fluctuations are suppressed. Thus one might surmise that the specific heat
will drop for large $\lambda$.  Typically,
the persistence length grows as $\xi_p \sim \exp(C\lambda)$; $C$ is inversely
proportional to the leading coefficient of the  beta function.  We would
expect that the peak position should shift  to the right with increasing $N$
in this scenario as
\begin{equation}
  \lambda_{peak}(N') - \lambda_{peak}(N) =
\frac{\ln(\frac{N'}{N})}{dC}.
\end{equation}
Quite a large value of $C$ is needed to explain the  rapid crossover; roughly
values of $C \sim 10, d_{intr} \sim 3$ are more or less consistent with the
magnitude of the peak shift and crossover width.  The RG calculations using
different forms of the action yield $C = \frac{4\pi}{3}$ (see equation
\ref{persist}), but
this may not apply to the action we simulate.
This reasoning also indicates that the peak should widen as the lattice size
increases; we do not observe this at all.
It seems plausible though that
these arguments, based only on the leading term of the high $\lambda$
expansion,
are too naive.

      An alternative scenario, which builds on the ideas in the above
paragraph, is suggested by the tantalizing similarities between the results of
our fluid surface simulations and what has been observed for the $d=4$ $SU(2)$
Lattice Gauge Theory \cite{ENZOHI} and for the $d=2$ $O(3)$ model.

The $O(3)$ model, which is thought to be
asymptotically free, exhibits a specific heat peak
near $\beta = 1.4$ (first measured via Monte Carlo
simulations by Colot \cite{COLOT}).  The
origin of this peak is understood \cite{BROUT}; it is due to the
fluctuations of the sigma particle, a low-mass bound state of the massless
$O(3)$ pions.  The sigma induces short-range order and contributes to the
specific heat as a degree of freedom only at high temperatures (when the
correlation length in the system becomes smaller than its inverse mass).
The peak thus occurs at the beginning of the crossover regime, when the
correlation length is several lattice spacings.

According to the low temperature expansion, the correlation  length grows as
$\xi \sim \exp(2\pi\beta)/\beta$. Thus one would expect  a fairly rapid
crossover in the $O(3)$ model;  the correlation length should increase by
roughly a factor of $9$ when $\beta$ is shifted by about $0.35$.
Such a crossover is indeed observed, though it
is not so apparent that it is as dramatic as the crossover behaviour
observed for fluid surfaces.
\footnote{To
compare quantitatively the width of the crossover regimes for these two models
it would
be necessary to measure a correlation length (perhaps
extracted from the normal-normal correlation function) in these random
surface simulations.}

Recent simulations of the $O(3)$ model \cite{KOSTAS} indicate that the
specific heat peak grows significantly when the lattice size $L$ is increased
from $5$ to $15$, and that virtually no growth in peak height is evident as
$L$ is increased further up to $100$.  Furthermore, the
peak position shifts to the
right as $L$ grows and then appears to stabilize for large $L$.  This is more
or less what we observe in our simulations of fluid surfaces, on lattices of
small size.  We point out these similarities largely to emphasize that there
does exist an asymptotically free theory (with low mass excitations) which
exhibits crossover behaviour qualitatively similar to that
observed in our simulations.
The analogy is perhaps deeper, though, since the fluid surface action (with
extrinsic curvature) in certain guises looks like a sigma model action.  It
would not therefore be so surprising from this point of view to find a sigma
particle in these theories, perhaps associated with ($\hat{n}^2 -1$), in which
$\hat{n}$ denotes the unit normal to our surfaces.

Another additional possibility is  that fluctuations of the intrinsic geometry
(the Liouville mode) are  responsible for short-range order and contribute to
the specific heat peak.

\section{Conclusion}
   We have introduced a model of fluid random surfaces with an extrinsic
curvature dependent action and explored its phase diagram.  Unfortunately,
we have been unable to determine
if our model undergoes a phase (crumpling) transition at finite coupling.
We have observed dramatic crossover behavior for particular observables
in our Monte Carlo simulations, but on the other hand, the correlation times
and certain finite-size effects do not behave as one would expect in the
presence of a phase transition.
The behavior of other lattice models also indicates that it is possible
that we are observing the effects of finite-mass excitations on small
lattices, rather than a phase transition.  We hope that future work
will clarify this murky state of affairs, to determine if there indeed exists
a crumpling transition for fluid surfaces.

\section{Acknowledgments}
This work has been done with NPAC (Northeast Parallel Architectures Center)
and CRPC (Center for Research in Parallel Computing) computing facilities.
The research of MB was supported by the Department of Energy Outstanding
Junior Investigator Grant DOE DE-FG02-85ER40231 and that of GH by
research funds from Syracuse University.  We gratefully acknowledge
discussions, help and sympathy from Jan Ambj{\o}rn, Kostas
Anagnostopoulos, John Apostolakis, Clive Baillie, Mike Douglas, David
Edelsohn, Geoffrey Fox, Volyosha Kazakov, Emil Martinec, Alexander
Migdal, David Nelson, Giorgio Parisi, Bengt Petersson, Steve Shenker, and
Roy Williams.  Deborah Jones, Peter Crockett, Mark Levinson and Nancy
McCracken provided invaluable computational support.

\vfill
\newpage
\section{Figure Captions}
  \begin{itemize}

    \item[Fig. 1] The specific heat $C(\lambda)$ as a function of $\lambda$.
As in all other pictures, filled circles and a dotted line correspond to
$N=144$, crosses and
a dashed line indicate $N=288$, and empty squares and a solid line
represent $N=576$.

    \item[Fig. 2a] The gyration radius $R_G$ defined in
(\ref{E_RG}), plotted as in Fig. 1.

    \item[Fig. 2b] The effective inverse Hausdorff dimension $\nu$ as a
function of $\lambda$, as defined in (\ref{E_RGD}). The filled dots and the
dashed curve are from a fit to the $N=288$ and $N=144$ data, while the empty
dots and solid curve represent the fit to $N=576$ and $N=288$.

    \item[Fig. 3] The extrinsic Gaussian curvature
$\mid{\cal{K}}\mid$ defined in (\ref{E_CK}), plotted as in Fig. 1.

    \item[Fig. 4] The  intrinsic curvature
$\mid{\cal{R}}\mid$ defined in (\ref{E_CR}), plotted as in Fig.1.

    \item[Fig. 5] The intrinsic extrinsic curvature
correlation, as defined in (\ref{E_KR}), plotted as in Fig.1.

    \item[Fig. 6a] A snapshot of a $576$ node surface of toroidal topology
for $\lambda = 0.8$.

    \item[Fig. 6b] As in Fig. 6a for $\lambda = 1.3$.

    \item[Fig. 6c] As in Fig. 6a for $\lambda = 1.5$.

    \item[Fig. 6d] As in Fig. 6a for $\lambda = 2.0$.
\end{itemize}
\vfill
\end{document}